\documentclass{optica-article}
\pdfoutput=1
\journal{opticajournal} % for journals or Optica Open

\articletype{Research Article}

\usepackage{graphicx}
\usepackage{float}
\usepackage{subfigure} 
\usepackage{lineno}
\usepackage{hyperref}
\hypersetup{hypertex=true,
colorlinks=true,
linkcolor=blue,
anchorcolor=blue,
citecolor=blue}
\definecolor{c1}{RGB}{39,82,201} % 目录颜色
\definecolor{c2}{RGB}{190,20,83} % 引用颜色
\hypersetup{
	colorlinks,
	linktoc = section, % 超链接位置，选项有section, page, all
	linkcolor = c1, % linkcolor 目录颜色
	citecolor = c1  % citecolor 引用颜色
}
\usepackage{cleveref}
\crefformat{figure}{#2{\textcolor{c2}{fig #1}}#3} % 图片的引用格式
\crefformat{equation}{#2{(\textcolor{c2}{#1})}#3} % 公式的引用格式
\crefformat{table}{#2{\textcolor{c2}{tab #1}}#3} % 表格的引用格式
\begin{document}

\title{Angle measurement method of electronic speckle interferometry based on Michelson interferometer }

\author{Siyuan Zhu,\authormark{1,*} Tao Li,\authormark{1} Zhongshan Chen,\authormark{1}and Xin Li\authormark{1}}

\address{\authormark{1}School of Physics and Astronomy, Sun Yat-Sen University, Zhuhai 519082, China}

\email{\authormark{*}zhusy29@mail2.sysu.edu.cn} %% email address is required; see note below about the corresponding author designation

% use {asbstract*} to suppress the copyright line. Copyright information will be added in production

%\address{3)}{}

%\cdate{2020-01-01}{2020-01-01}

\abstract{This paper proposes an angle measurement method based on Electronic Speckle Pattern Interferometry (ESPI) using a Michelson interferometer. By leveraging different principles within the same device, this method achieves complementary advantages across various angle ranges, enhancing measurement accuracy while maintaining high robustness. By utilizing CCD to record light field information in real time and combining geometric and ESPI methods, relationships between small angles and light field information are established, allowing for the design of relevant algorithms for real-time angle measurement. Numerical simulations and experiments were conducted to validate the feasibility and practicality of this method. Results indicate that it maintains measurement accuracy while offering a wide angle measurement range, effectively addressing the limitations of small angle measurements in larger ranges, showcasing significant potential for widespread applications in related fields.}
\section{Introdution}
As a crucial aspect of precision measurement research, small angle measurement is widely applied in fields such as precision machining, aerospace, and system calibration\cite{zhong2008absolute,moon2014truncated,Bin2015based,kumar2020technologies}. For high-precision detection of small angles, measurement errors are typically required to be at the arcsecond level or lower. Traditional mechanical and electromagnetic angle measurement methods often rely on manual and contact techniques, which cannot meet the demands for high-precision small angle measurements. In contrast, optical-based angle measurement methods\cite{Zhou2022fp} can achieve non-contact, high-precision, and high-resolution angle measurements, leading to their increasing application in the field of small angle measurement in recent years.

Optical angle measurement methods can be broadly categorized into monocular vision methods\cite{dong2015practical,li2010method}, interferometry\cite{shi2018roll,zhang2014laser}, polarization change methods\cite{qi2019heterodyne,gillmer2015robust}, parallel beam methods\cite{cai2019robust,ren2020parallel,fan2021self}, and self-collimation methods\cite{chen2017optical,zhu2013common, li2022roll}. Although these techniques have been effectively applied in scientific research and engineering, they still face challenges such as complex optical paths, poor stability, and limited measurement spaces\cite{wu2017high}. In recent years, the introduction of Electronic Speckle Pattern Interferometry (ESPI) has emerged as a high-precision, full-field optical interference technique well-suited for directly measuring in-plane components\cite{wu2019precision}. ESPI effectively mitigates the effects of stray light interference that often plague traditional interferometric systems, thereby enhancing system stability. Similar research models have been reported, demonstrating that ESPI is an effective tool for angle measurement\cite{wu2021simultaneous,wang2018temporal}. However, ESPI has limitations regarding its measurement range, and its measurement precision primarily relies on the accuracy of phase extraction. Therefore, designing efficient and accurate phase extraction techniques and expanding the measurement range remain significant challenges.

This paper proposes a novel angle measurement method based on Electronic Speckle Pattern Interferometry (ESPI) using a Michelson interferometer. By employing different angle measurement methods across various angle ranges, it achieves high precision, high resolution, and a wide measurement range. Specifically, the geometric method records the movement of the centroid of interference spots through two exposures, inferring small angles from the relationship between angle and centroid movement, benefiting from a wide measurement range and strong interference resistance. The ESPI method captures phase changes across angles in two exposures, deducing small angles from the phase variation while simplifying traditional phase extraction algorithms and designing more precise methods. The second section will discuss the theoretical derivation and numerical simulations of both methods, while the third section will present experimental results and error analysis, with conclusions provided in the fourth section.

\section{Measurement Theories and Simulation Analysis}
\subsection{Theories}

Based on the experimental setup, an equivalent optical path is constructed (as shown in Fig 2.1). In a standard equal-inclination interference setup, where the mirrors are perpendicular to the incident light, the optical path does not deviate. The virtual light sources $S_{1}$, $S_{2}$ and the center $P_{0}$ of the screen are colinear. Under the circumstances, a series of concentric rings formed around $P_{0}$ in the interference pattern. In this scenario, if the mirrors $M_{1}$ and $M_{2}$ are slightly tilted within the $Oxz$ plane as depicted in Fig 2.1, resulting in small deviations $\theta_{1}$ and $\theta_{2}$ respectively, the virtual light sources $S_{1}$ and $S_{2}$ will move along an arc path.

By extending the connection line between S1 and S2, the intersection point $P_{c}$ is obtained as a new interference point center. Based on the known angles and the lengths of the optical arms, we can deduce the length of $P_{c}P_{0}$. In the context of discussing small angle changes, we can approximate $cos2(\theta_{2}-\theta_{1})\approx1$ and $sin2(\theta_{2}-\theta_{1})\approx2(\theta_{2}-\theta_{1})$. Combining these conditions, we obtain:
\begin{equation}\label{1}
    P_{c}P_{0}=\frac{2(d_{1}+2d_{3}+d_{4})(d_{1}+d_{2})(\theta_{2}-\theta_{1})}{2d_{2}-2d_{3}}
\end{equation}

Eqs.\eqref{1} indicates that $P_{c} P_{0}$ is proportional to $(\theta_{2} - \theta_{1})$. If we fix the mirror $M_{1}$, according to this relationship, we can determine the magnitude of the deviation angle of $M_{2}$ by observing the displacement of the interference fringes before and after the shift. Let $\delta \theta_{2}$ denote the deviation angle of $M_{2}$ before and after the shift, and $\Delta R$ denote the displacement of the interference fringes. The corresponding relationship is given by:
\begin{equation}\label{2}
 \Delta R=\frac{2(d_{1}+2d_{3}+d_{4})(d_{1}+d_{2})}{2d_{2}-2d_{3}}\cdot\delta\theta_{2}
\end{equation}

\begin{figure}[htbp]
\renewcommand{\thefigure}{2.1}
\centering\includegraphics[width=7.5cm]{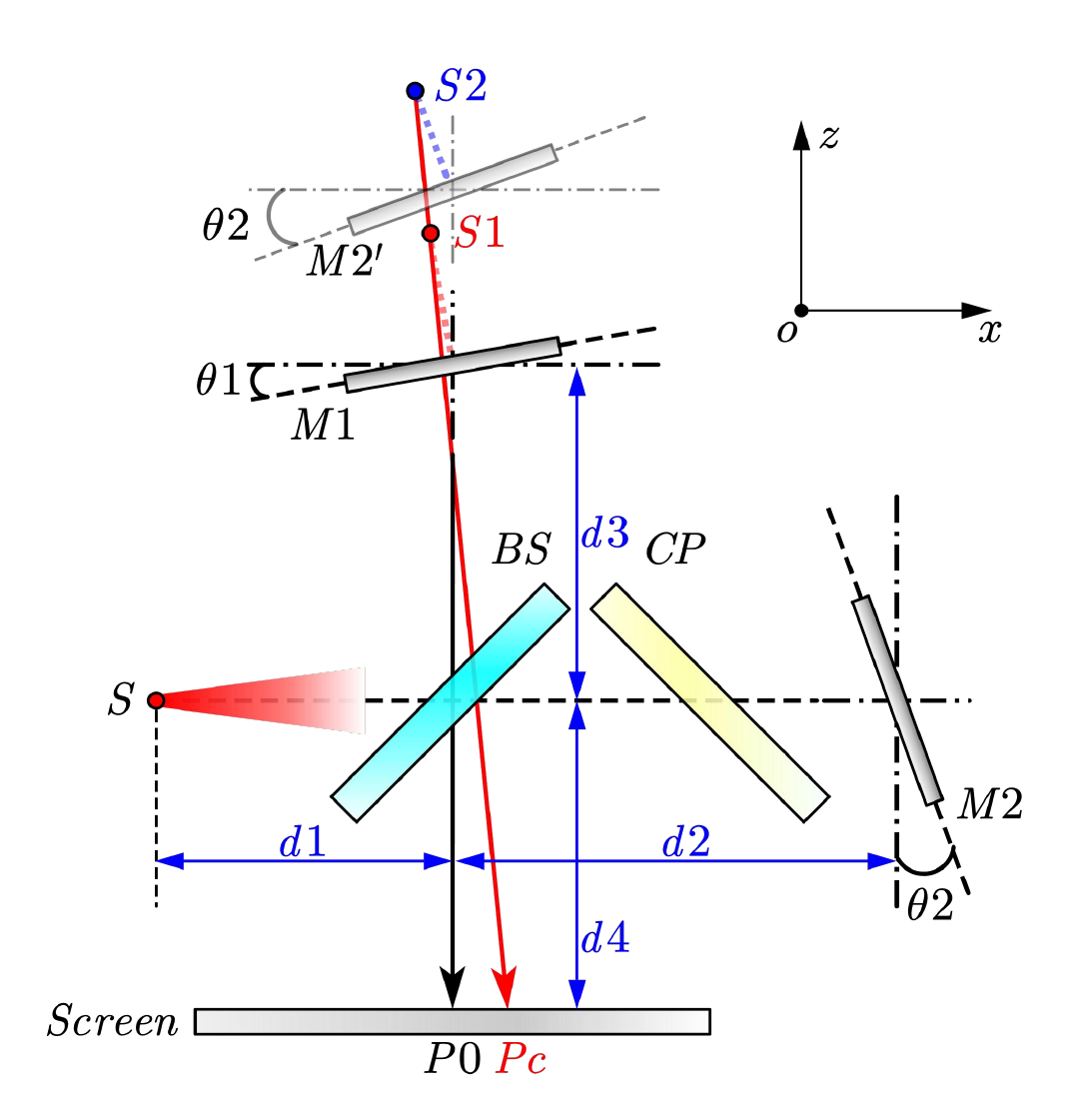}
\caption{The principle diagram of the experimental setup's equivalent optical path is shown below, where BS represents the beam splitter, CP is the compensator, and M1 and M2 are both rotatable plane mirrors. The optical path corresponds to the Oxz plane, while the screen corresponds to the Oxy plane. In the diagram, $S_1$ and $S_2$ are equivalent virtual light sources.}
\end{figure}

\hspace*{\fill} \

Similar to the general ESPI method, interference fringes can be regarded as the interference between two beams of light emitted from $S_{1}$ and $S_{2}$ on the screen. After passing through the beam expander, the light emitted from the virtual light sources can be considered as spherical waves. Assume that the wavefront of the light emitted from point $S_{1}$ is $U_{1}=a_{1}e^{i\phi_{1}}$, and the wavefront of the light emitted from point $S_{2}$ is $U_{2}=a_{2}e^{i\phi_{2}}$, where $\phi_{1}$ and $\phi_{2}$ represent the phase distributions of $S_{1}$ and $S_{2}$ on the screen respectively. The intensity distribution after superposition is given by the following equation.
\begin{equation}\label{3}
    I=(U_{1}+U_{2})(U_{1}+U_{2})^{\ast}=a_{1}^{2}+a_{2}^{2}+2a_{1}a_{2}\cos(\phi_{2}-\phi_{1})
\end{equation}

To simplify Eqs.\eqref{3}, we denote the background light as $I_{0}=a_{1}^{2}+a_{2}^{2}$, the phase as $\phi=\phi_{2}-\phi_{1}$, the contrast as $\gamma=\frac{2a_{1}a_{2}}{a_{1}^{2}+a_{2}^{2}}$. Subsequently, the intensity can be expressed as $I=I_{0}(1+\gamma\cos\Delta)$.Here, we use $\delta \phi$ to represent the change in the phase distribution of the intensity on the screen, which is caused by the change in angle generated by mirror . Thus, the intensity can be described as follows: 
\begin{equation}
    I'=I_{0}[1+\gamma\cos(\phi+\delta \phi)]
\end{equation}

According to this relationship, when the angle of mirror M2 is changed, We can subtract the intensity distributions before and after the change, and then utilize the corresponding processing to obtain the phase change caused by the angle change. The variation of the optical field is determined by the following equation:

\begin{equation}
    \Delta I=|I'-I|=\left|I_{0}\gamma[\cos(\phi+\delta \phi)-\cos\phi]\right|=\left|2I_{0}\gamma\sin\left({\frac{\delta \phi}{2}}\right)\sin\left(\phi+{\frac{\delta \phi}{2}}\right)\right|
\end{equation}

Now, consider that mirror $M2$ undergoes a small angular deviation in the $Oxz$ plane, without considering the deviation in the y-direction. Under the condition of small angles, we can approximate $cos\theta_{2}\approx1$ and $sin\theta_{2}\approx\theta_{2}$. After the small angular deviation $\theta_{2}$, denoted as $\theta_{2}'$, the screen is located in the $Oxy$ plane in the coordinate system of the diagram. Let $(x, y)$ denote the coordinates of any point on the screen. We can calculate the optical path length for any point and simplify the calculation of the optical path difference before and after the deviation for each point.
\begin{equation}
    \Delta\mathrm{L}=(d_{1}+2d_{2}+d_{4})\Biggl\{\sqrt{1+{\frac{[x+(d_{1}+d_{2})2\theta_{2}^{\prime}]^{2}+y^{2}}{[d_{1}+2d_{2}+d_{4}]^{2}}}}-\sqrt{1+{\frac{[x+(d_{1}+d_{2})2\theta_{2}]^{2}+y^{2}}{[d_{1}+2d_{2}+d_{4}]^{2}}}}\Biggr\}
\end{equation}

According to the Taylor series expansion form $\sqrt{1+x}=1 + \frac{1}{2}x - \frac{1}{8}x^{2} + O(x^{3})$, we simplify the expression for the optical path difference, retaining only the first-order term. Then, we convert the optical path difference into a phase difference to obtain:
\begin{equation}\label{7}
   \delta \phi=\frac{2\pi}{\lambda}\frac{2\theta(d_{1}+d_{2})}{(d_{1}+2d_{2}+d_{4})}x
\end{equation}

$\theta = \theta_{2}' - \theta_{2}$ represents the deviation angle, $\lambda$ is the wavelength of light, and according to Eqs. \eqref{7}, the phase difference $\delta = n\pi (n=0,1,2...)$ produces a minimum in the intensity information. Therefore, parallel fringes with uniform spacing will appear along a straight line in the optical field. From the minimum condition, we obtain the expression for the fringe spacing as follows:
\begin{equation}\label{8}
    d=\frac{\lambda(d_{1}+2d_{2}+d_{4})}{2\theta(d_{1}+d_{2})}
\end{equation}

To measure the fringe spacing $d$, substitute Eqs.\eqref{8} to obtain the magnitude of the deflection angle. Considering the 
$y$-direction equivalent to the $x$-direction, the rotation is only discussed on the $Oxy$ plane. In practical applications, the reflector can rotate in three dimensions: 
$Oxz$, $Oyz$, and$Oxy$ planes, corresponding to three types of rotations: yaw, pitch, and roll, respectively. Yaw and pitch are equivalent, as explained earlier, while roll will change the orientation of the fringes. By comparing the fringe orientation before and after rolling, the specific value of the rolling angle can be calculated. Describing the fringes with a straight line, the initial orientation is $y = tan\theta_{1}x$, and after the rolling angle, it becomes $y = tan\theta_{2}x$, Therefore, the rolling angle can be obtained as $\theta_{2} - \theta_{1}$.
\\~
\subsection{Simulations}

Using Zemax and Matlab simulations respectively, the setup and optical path are illustrated in Fig 2.1. Adjust the tilt angle of the mirror $M_2$ and sequentially save images to obtain optical field data at different angles. The simulated images obtained from both methods are shown in Fig 2.2.

\begin{figure}[htbp]
\renewcommand{\thefigure}{2.2}
\centering
\subfigure[Zemax:0'']{
\label{time.1}
\includegraphics[width=2.2cm]{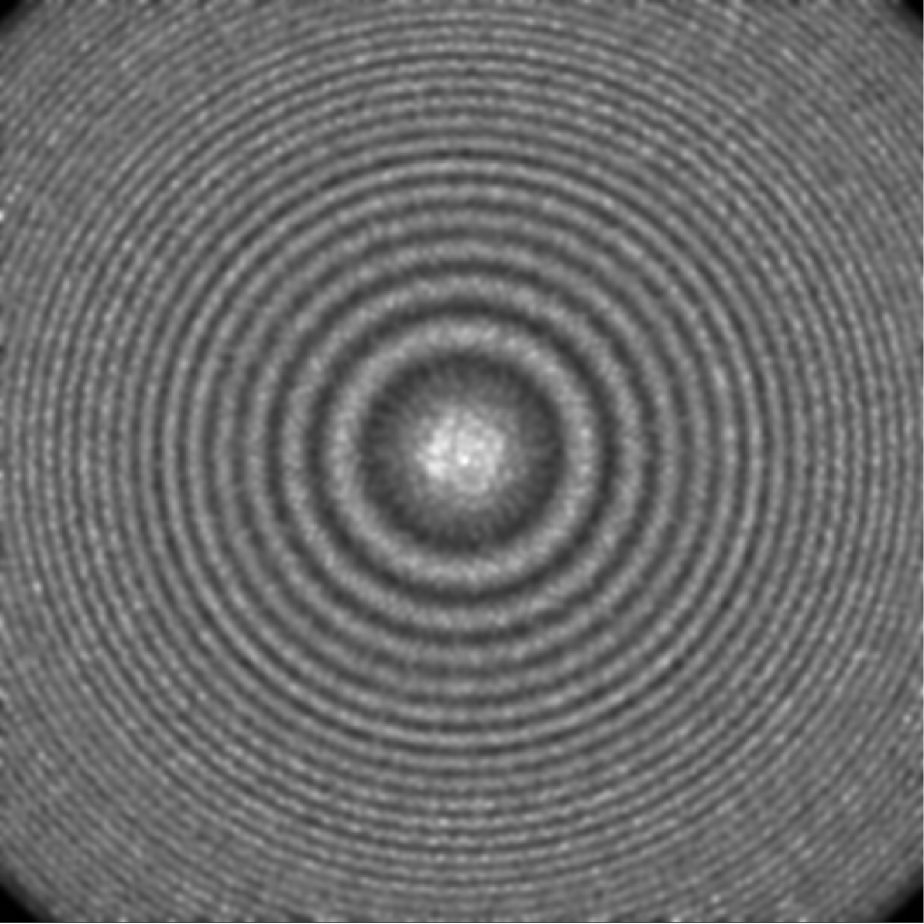}
}
\subfigure[Zemax:33'']{
\label{time.2}
\includegraphics[width=2.2cm]{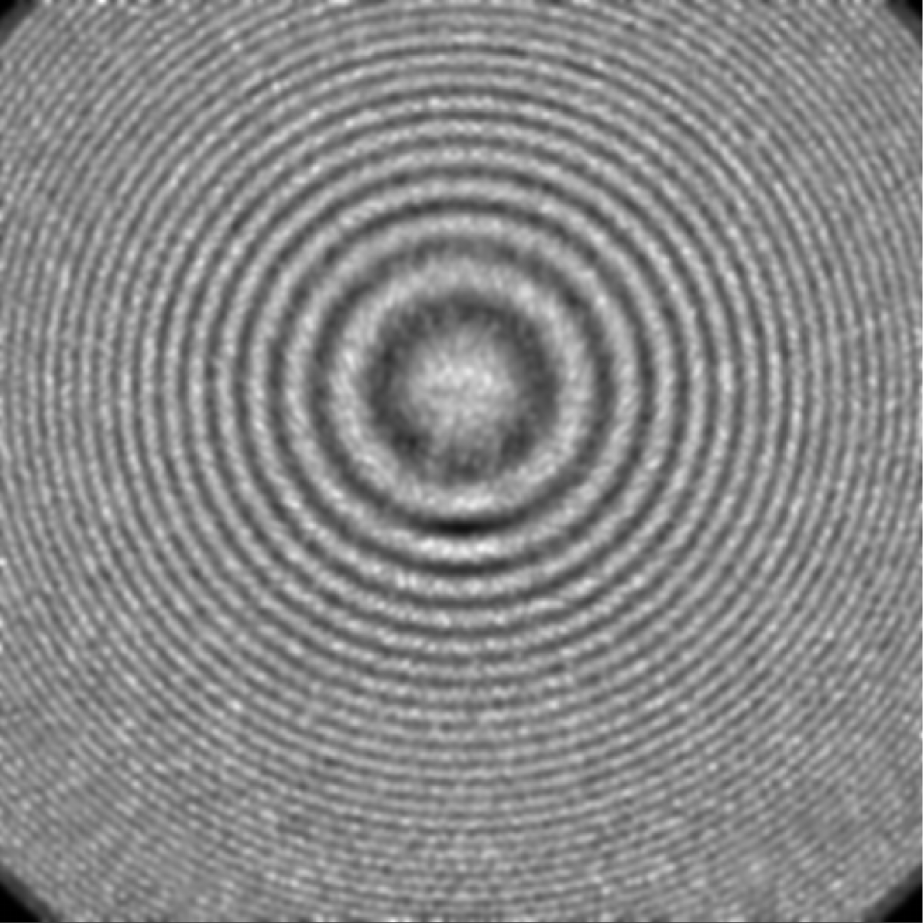}
}
\subfigure[Zemax:66'']{
\label{time.3}
\includegraphics[width=2.2cm]{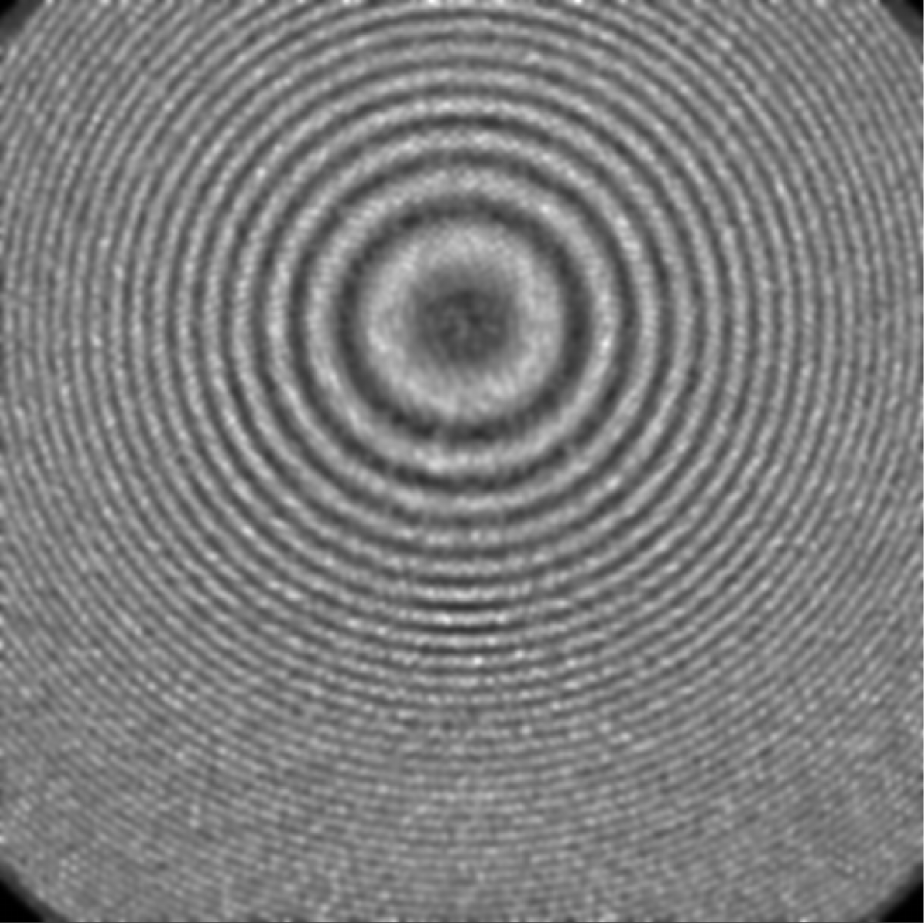}
}
\\
\subfigure[Matlab:0'']{
\label{time.4}
\includegraphics[width=2.2cm]{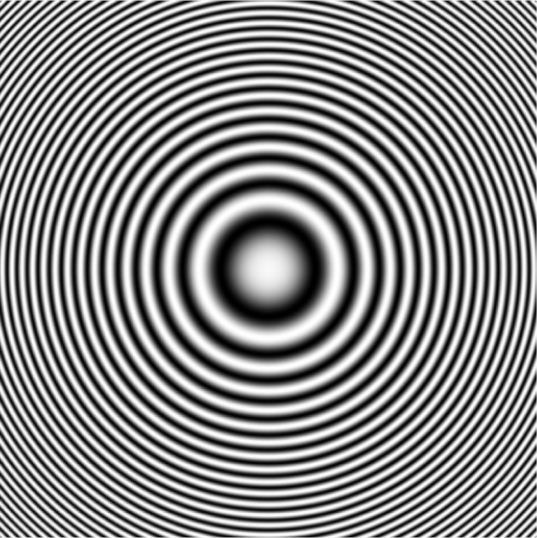}
}
\subfigure[Matlab:33'']{
\label{time.5}
\includegraphics[width=2.2cm]{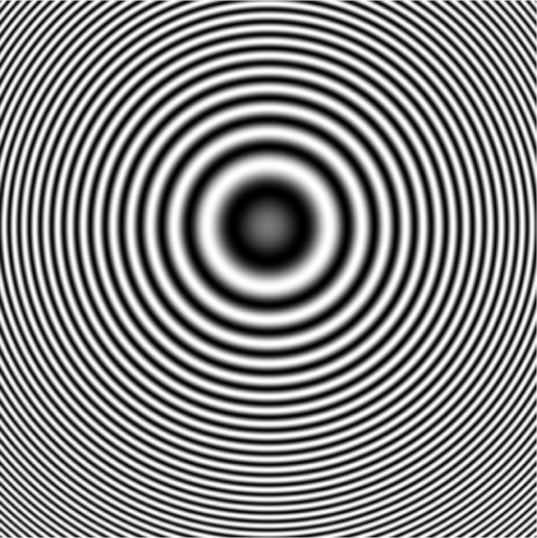}
}
\subfigure[Matlab:66'']{
\label{time.6}
\includegraphics[width=2.2cm]{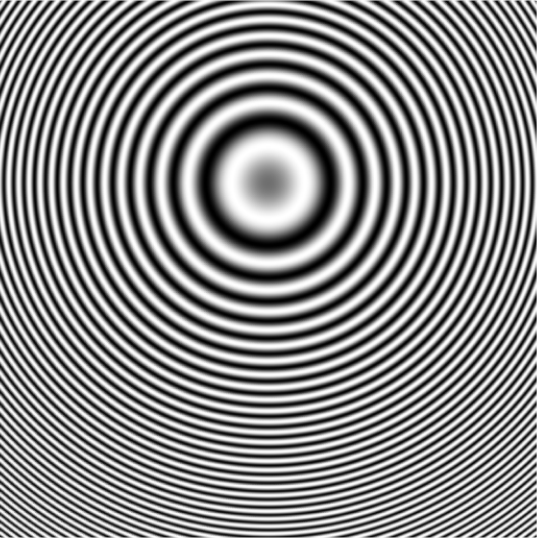}
}
\caption{Zemax and Matlab simulation images}
\label{fig:几何仿真}
\end{figure}

According to the theory of geometric optics, there exists a linear relationship between the interference center and the deflection angle. The centroid tracking method is used to obtain the center coordinates of the circular spot, and displacement is calculated from this. By using Eqs.\eqref{2} to back-calculate angle data from the measured displacement data, the relationship between the set angle of the mirror and the measured angle from simulation can be obtained, as shown in Fig 2.3. The maximum relative error between the set angle and the measured angle is 1.75\%, with an overall average error of -0.35\%, which validates Eqs.\eqref{2}.

\begin{figure}[htbp]
\renewcommand{\thefigure}{2.3}
\centering
\subfigure[]{
\label{time.1}
\includegraphics[width=6cm]{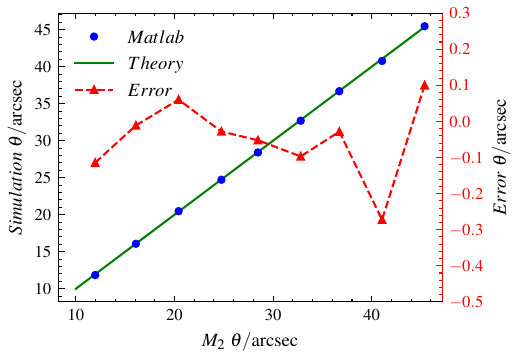}
}
\subfigure[]{
\label{time.2}
\includegraphics[width=6cm]{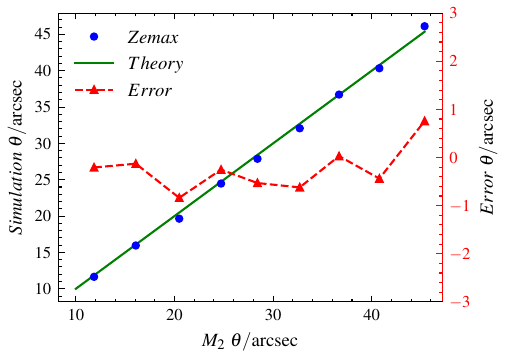}
}
\caption{The simulation results of geometric method. (a) Matlab simulation,(b) Zemax simulation. During simulation, the arm lengths are set to $d_1=7.66cm$, $d_2=17.9cm$, $d_3=17.2cm$, and $d_4=13.2cm$. The green line represents the theoretical curve $y=x$, while the red dashed line represents the absolute error of the simulation.}
\label{图1的标签}
\end{figure}

ESPI method requires preprocessing of the images by subtracting the light intensity information of the two images. According to the theory of ESPI, the processed image will exhibit evenly spaced fringes in the same direction, and the fringe spacing is determined by Eqs.\eqref{8}. The processed image is shown in Fig 2.4.

When the optical arm lengths $d_{1}$, $d_{2}$, $d_{3}$, and $d_{4}$ remain constant, it can be visually observed from Fig 2.4 that as the angular difference $\theta$ increases, the spacing between fringes decreases. This is consistent with Eqs.\eqref{8}, which states that the fringe spacing is inversely proportional to $\theta$. By measuring the fringe spacing and substituting it into the formula, the relationship between the set angle of the reflector and the simulated measurement angle can be obtained. As shown in Fig 2.5, the maximum relative error between the set angle and the measured angle is -0.95\%, and the overall average error is -0.32\%, which validates Eqs.\eqref{8}.
\begin{figure}[htbp]
\renewcommand{\thefigure}{2.4}
\centering
\vspace{0.2cm}
\subfigure[Zemax:16.5"]{
\label{time.1}
\includegraphics[width=2.2cm]{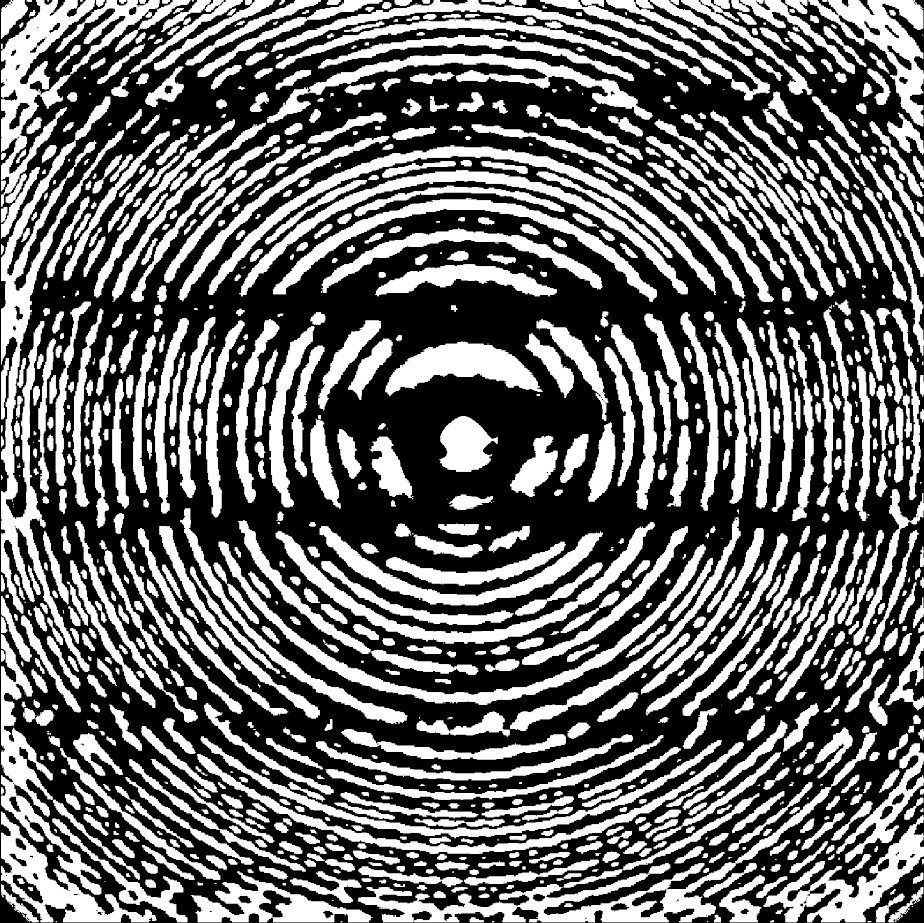}
}
\subfigure[Zemax:33"]{
\label{time.2}
\includegraphics[width=2.2cm]{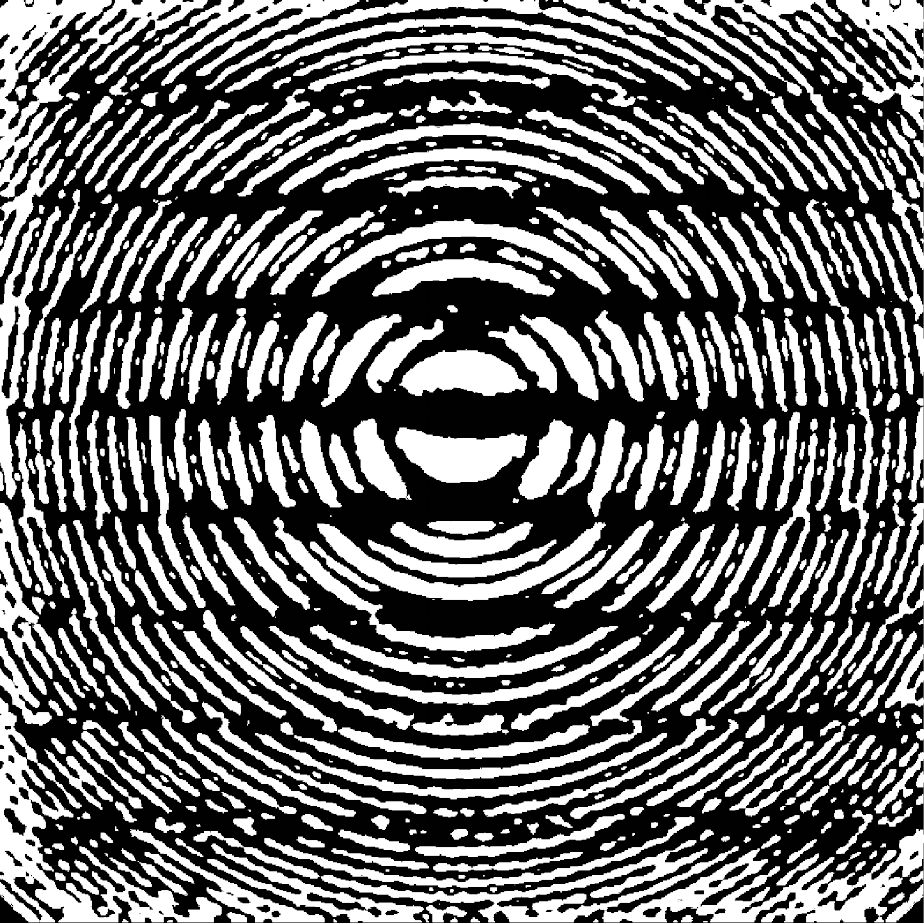}
}
\subfigure[Zemax:49.5"]{
\label{time.3}
\includegraphics[width=2.2cm]{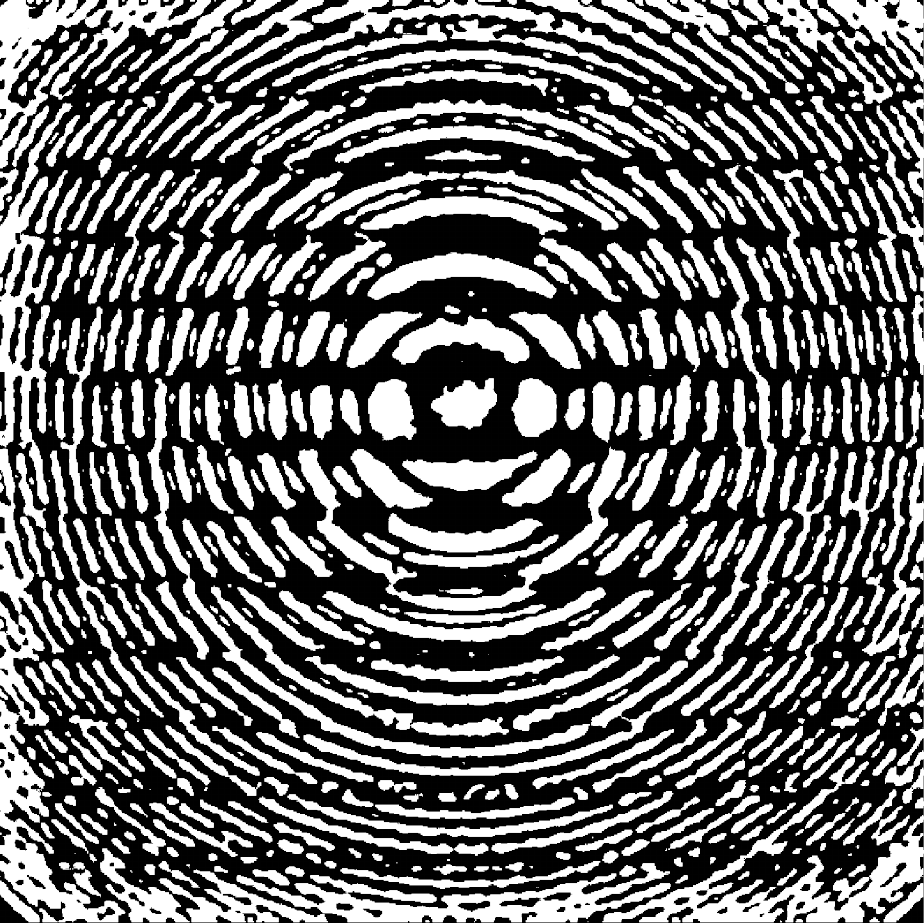}
}
\\
\subfigure[Matlab:16.5"]{
\label{time.4}
\includegraphics[width=2.2cm]{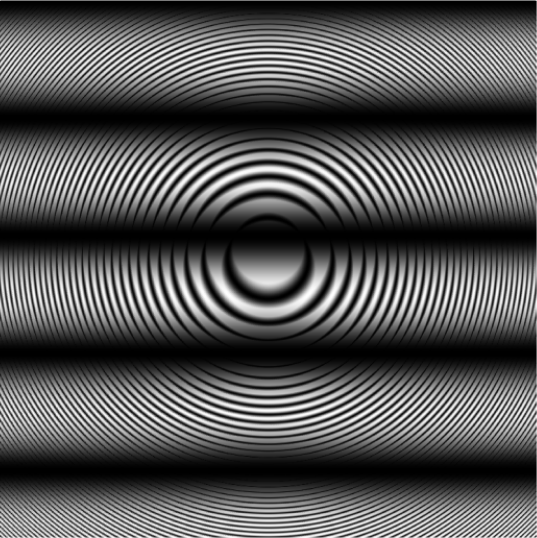}
}
\subfigure[Matlab:33"]{
\label{time.5}
\includegraphics[width=2.2cm]{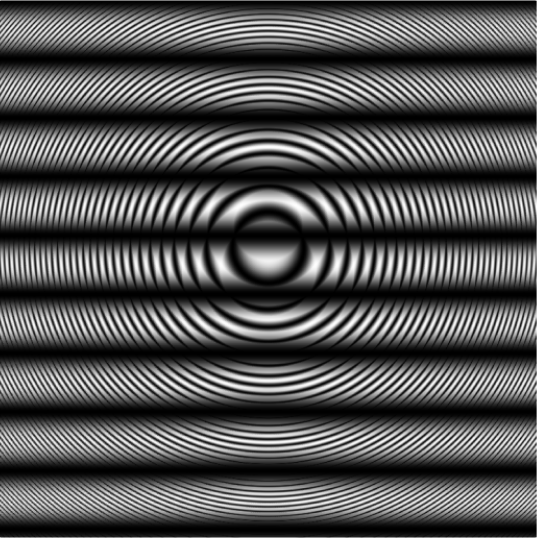}
}
\subfigure[Matlab:49.5"]{
\label{time.6}
\includegraphics[width=2.2cm]{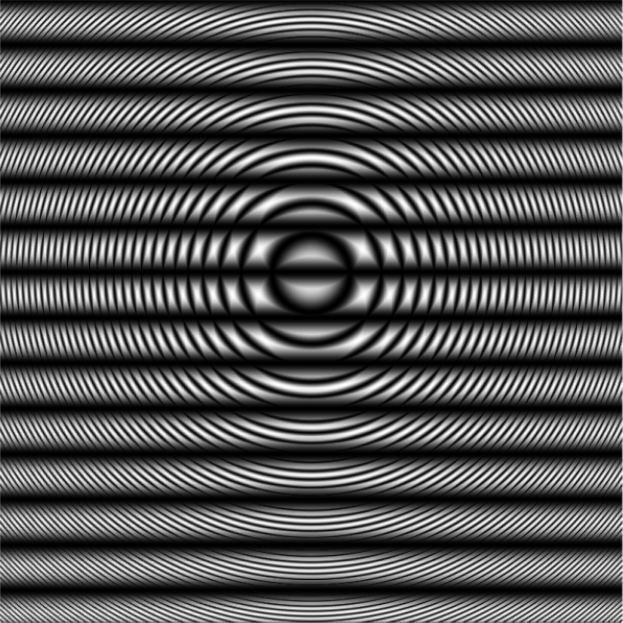}
}
\caption{The processed interference images from Zemax and Matlab simulations exhibit evenly spaced interference fringes.}
\label{图1的标签}
\end{figure}

\begin{figure}[htbp]
\renewcommand{\thefigure}{2.5}
\centering
\subfigure[]{
\label{time.1}
\includegraphics[width=6cm]{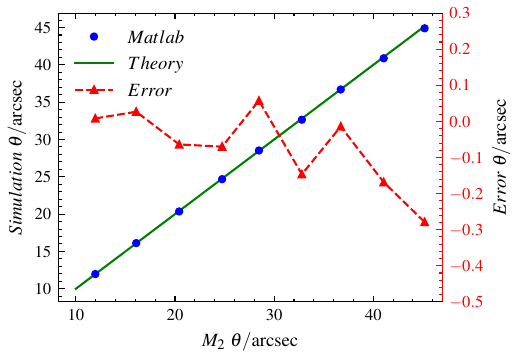}
}
\subfigure[]{
\label{time.2}
\includegraphics[width=6cm]{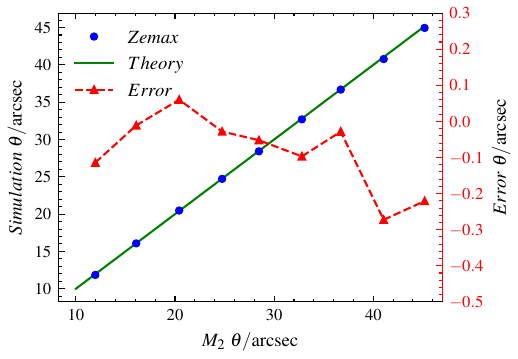}
}
\caption{ESPI method simulation results are shown in (a) for Matlab simulation and (b) for Zemax simulation. During the simulation, the arm lengths are set to $d_1=7.66cm$, $d_2=17.9cm$, $d_3=17.2cm$, and $d_4=13.2cm$. The green line represents the theoretical curve $y=x$, while the red dashed line represents the absolute error of the simulation.}
\label{图1的标签}
\end{figure}

As the measured deflection angle increases, according to Eqs.\eqref{2} and \eqref{8}, to ensure the high precision of the measurement results, it is necessary to adjust the optical arm lengths appropriately to ensure that the $\Delta R$ and $d$ measured by both geometric and ESPI methods are within the field of view. Based on the simulation results, as shown in Fig 2.6, by adjusting the optical arm lengths, within a range of 0.6°, the maximum relative error of the Angle measured by the geometric method is 1.03\% and the overall average error is 0.76\% within the range of 0.6°. The maximum relative error of the ESPI method is -0.08\%, and the average error of ESPI is -0.07\%.

However, the ESPI method requires the ability to resolve a spacing of 0.0669mm within a range of 0.6°, CCDs in experiments often cannot achieve such high resolution. In contrast, the geometric method is not affected by this limitation. Therefore, the geometric method is used to achieve more accurate angle measurement in the larger angle, and in the extremely angle range of 0.015°, the ESPI method will be more advantageous.

\begin{figure}[htbp]
\renewcommand{\thefigure}{2.6}
\centering
\subfigure[]{
\label{time.1}
\includegraphics[width=6cm]{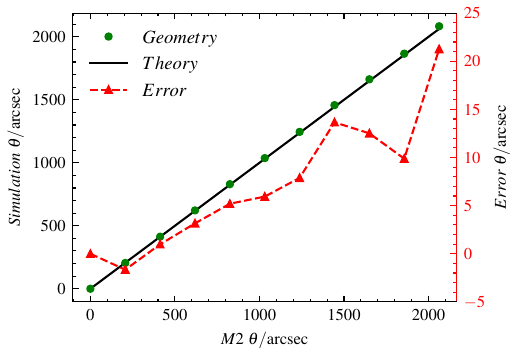}
}
\subfigure[]{
\label{time.2}
\includegraphics[width=6.11cm]{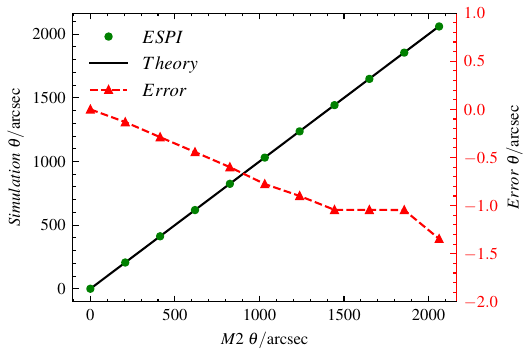}
}
\caption{The simulation results for large-angle measurements in Matlab  (a) for geometric method simulation and (b) for ESPI method simulation. During the simulation, the arm lengths are set to $d_1=7.2cm$, $d_2=17.85cm$, $d_3=18.2cm$, and $d_4=10.1cm$. The green line represents the theoretical curve $y=x$, while the red dashed line represents the absolute error of the simulation.}
\label{图1的标签}
\end{figure}

\subsection{Measurement}
The CCD recorded the interference images before and after the deflection mirror's adjustments through multiple exposures. After the experiment, image recognition algorithms are used to obtain the center position of the interference field and the light field distribution information, corresponding to the two methods proposed earlier. The images processing results for the two methods are shown in Fig 2.7.

\begin{figure}[htbp]
\renewcommand{\thefigure}{2.7}
\centering
\subfigure[]{
\label{time.1}
\includegraphics[width=3.5cm]{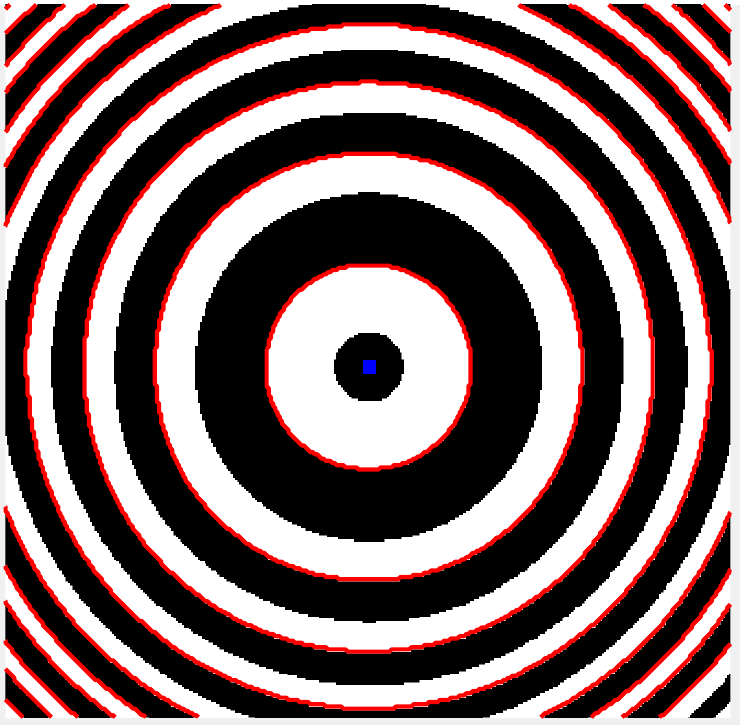}
}
\subfigure[]{
\label{time.2}
\includegraphics[width=3.5cm]{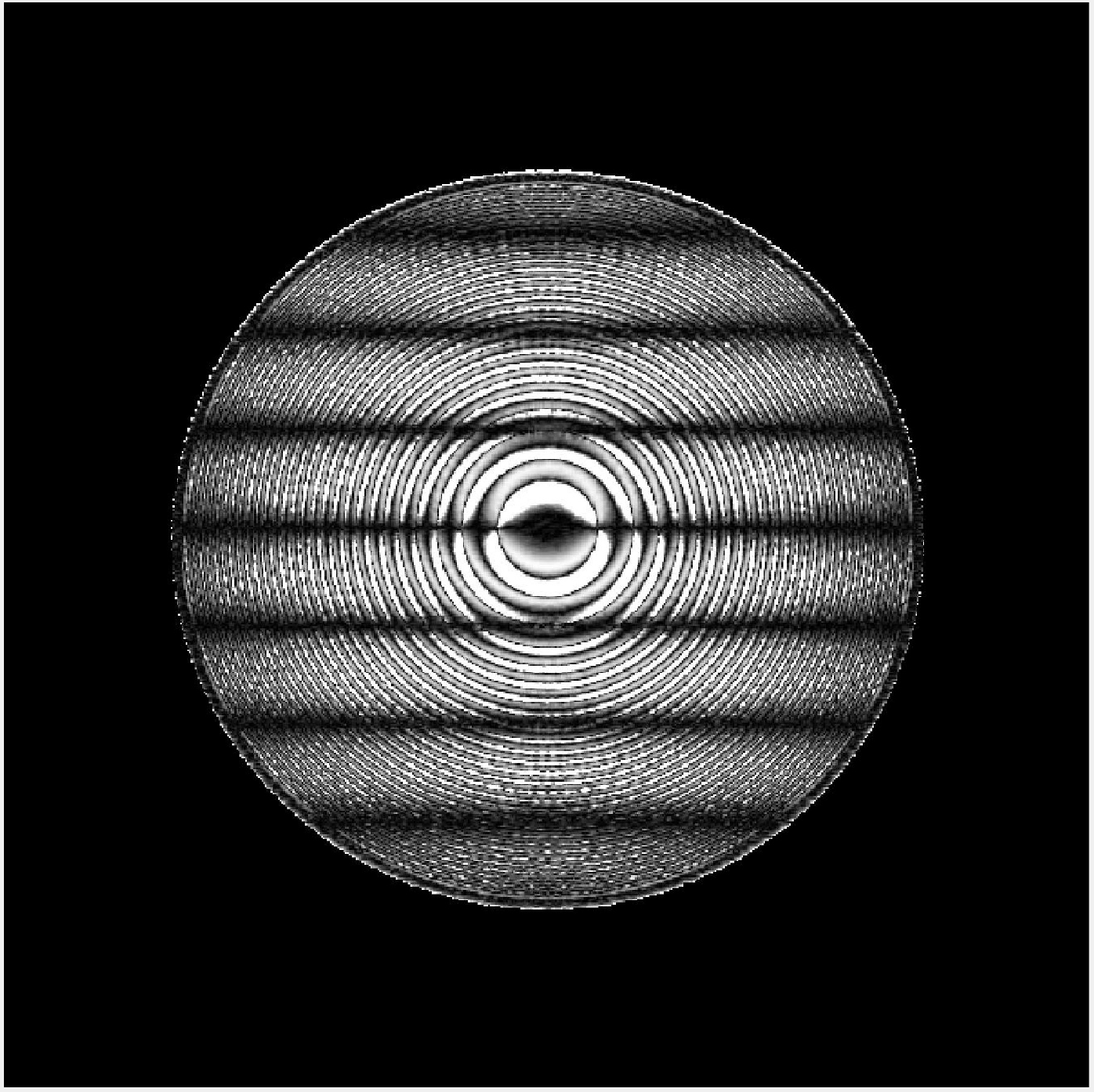}
}
\subfigure[]{
\label{time.3}
\includegraphics[width=3.5cm]{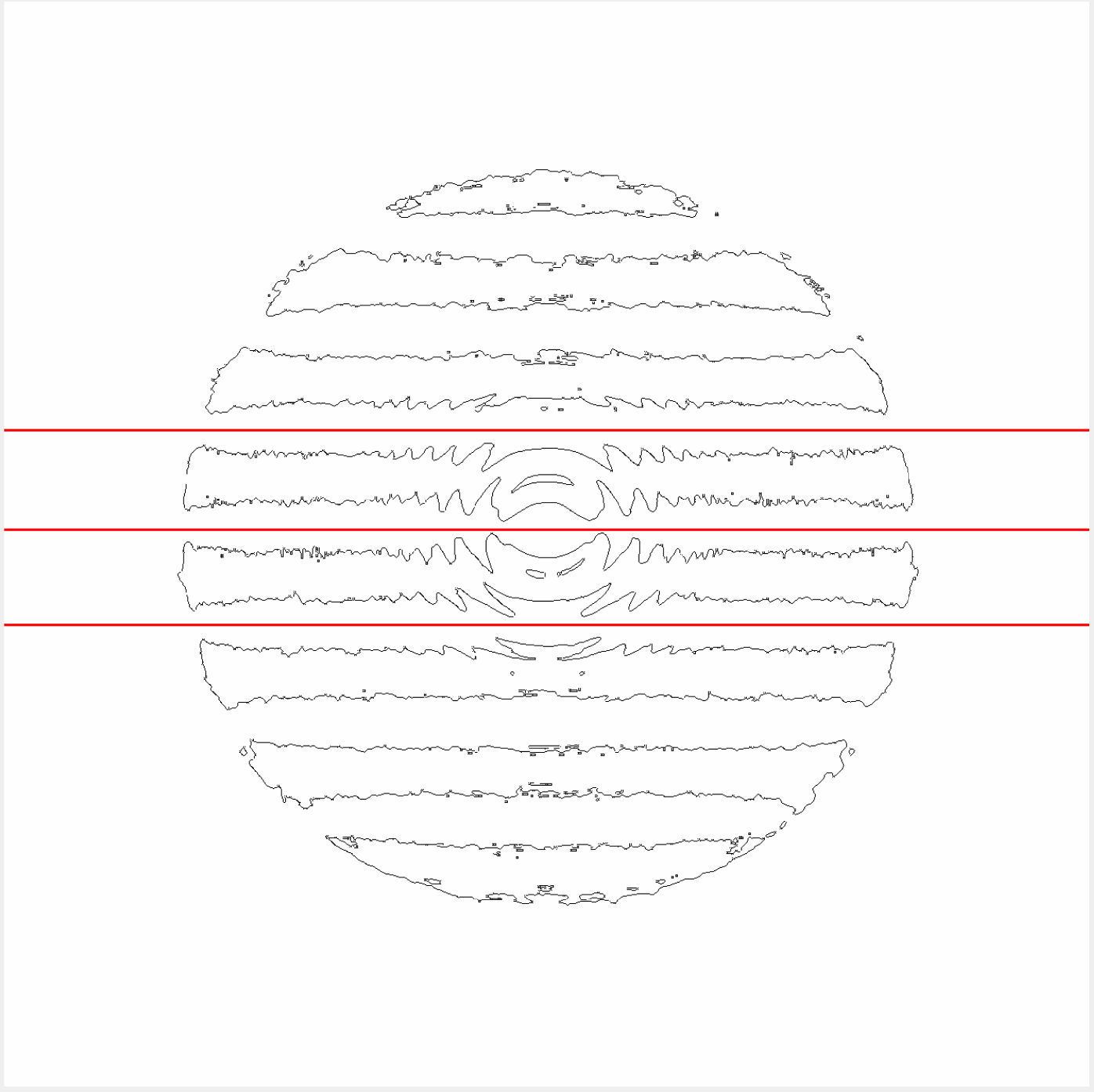}
}
\caption{(a) The extraction method for the geometric method's circle center coordinates involves obtaining the coordinates of the circle center ($x_i, y_i$) by extracting the edge of the ring (red circular line);(b) The image processed by the ESPI program, showing new evenly spaced interference fringes;(c) The result of binarizing the image in (b) and then extracting the edges. The red line represents the straight line fitted to the fringes using the least squares method, which is then used to determine the spacing of each fringe.}
\label{图1的标签}
\end{figure}

In processing the geometric method, interference images recorded at different deflection angles $\theta$ of the deflection mirror (M2) are imported into the MATLAB program. As shown in Fig 2.7(a), the images are binarized and noise points are removed. Then, the edges of the images are extracted to determine the center positions ($x_i, y_i$) of the interference rings. Eqs.\eqref{2} indicates that the movement of the center position is determined by the deflection angle $\theta$ of M2 and the lengths of the optical arms. Therefore, by substituting ($x_i, y_i$) into Eqs.\eqref{2}, the deflection angle $\theta_{Geo}$ can be calculated. For the ESPI method, the initial interference image and the interference image at the selected deflection angle $\theta$ of M2 are imported into MATLAB. The grayscale matrices of these images are subtracted, as shown in Fig 2.7(b). The resulting image from the subtraction process reveals new evenly spaced interference fringes. The direction of the interference fringes is determined by the deflection direction of the deflection mirror, and the fringe spacing is determined by the lengths of the optical arms and the deflection angle $\theta$ of the deflection mirror. As shown in Fig 2.7(c), the new interference image is binarized and edge extraction is performed. Then, the fringe lines in the image are fitted using the least squares method to obtain the fringe spacing $d$. By substituting $d$ into Eqs.\eqref{8}, the deflection angle $\theta_{ESPI}$ can be calculated.

~\\
\section{Experiment}

\subsection{Experimental Setup}
The experimental setup as depicted in Fig 3.1. The experimental platform consists of the following components: a 632.8nm He-Ne laser DH-HN300 model from Daheng Optics, a 6.2mm focal length beam expander, a mirror, a beam splitter, a compensating plate, a CCD camera with a resolution of 1920×1080 pixels, a ground glass receiver screen, and a piezoelectric deflection mirror assembly from Xinmingtian, model S34.T4SY. The specific layout is shown in Fig 3.3. The mirror is a flat mirror with adjustable angles, and the deflection mirror can generate a minimum deflection angle of 4.12". The relative positions of each component are fixed, with only the flat mirror adjustable in the forward and backward positions using a spiral micrometer with a resolution of 0.01mm.

\begin{figure}[htbp]
\renewcommand{\thefigure}{3.1}
\centering
\includegraphics[width=6cm]{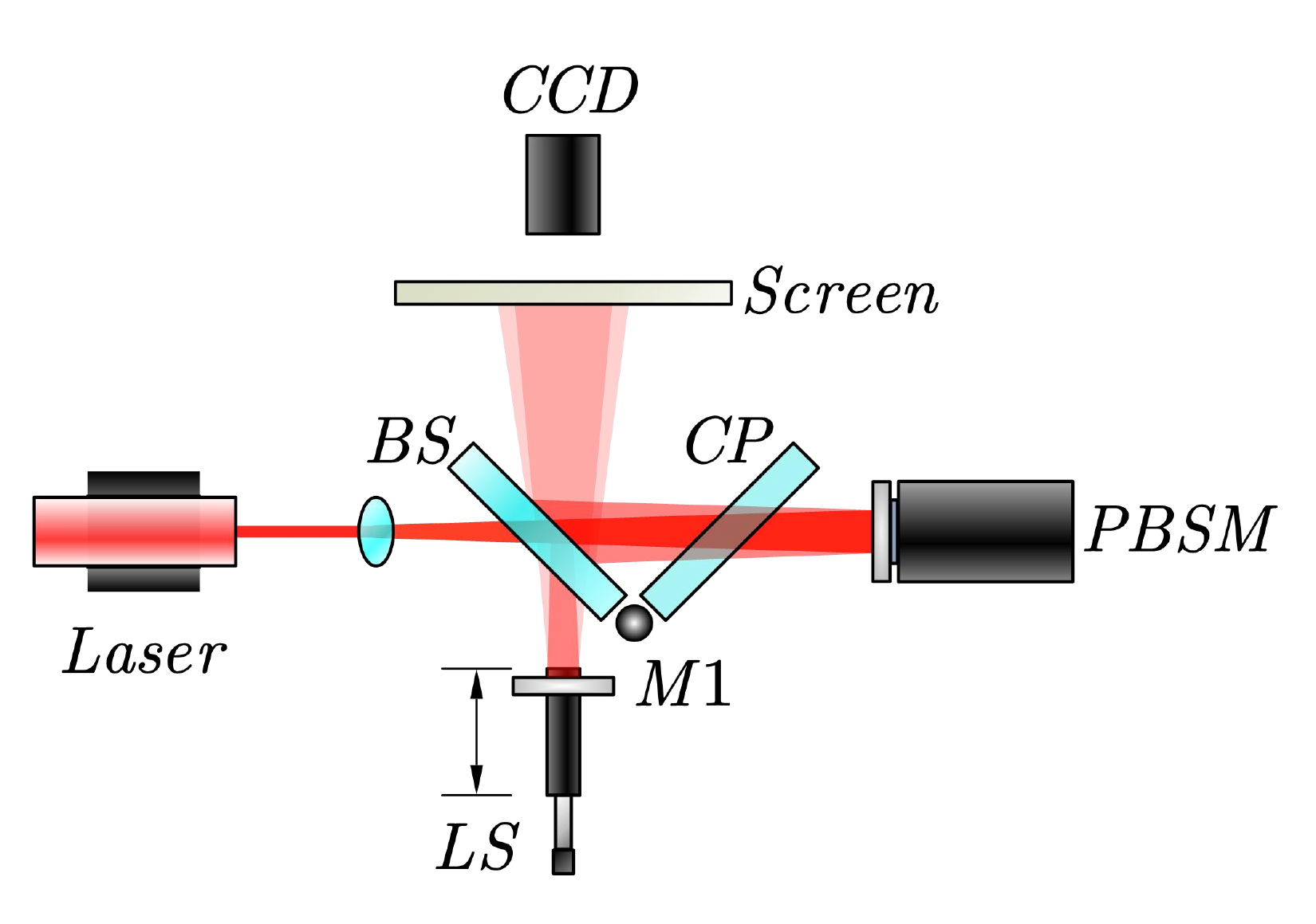}
\caption{The schematic diagram of the experimental setup.BS$:$ beam splitter, CP$:$ compensating plate, PBSM$:$ piezoelectric deflection mirror, CCD$:$ camera, LS$:$ the movable platform adjusted by a spiral micrometer. The plane of the ground glass receiver screen corresponds to the Oxy plane, and the optical path diagram corresponds to the Oxz plane.}
\end{figure}

The collimated laser is emitted from the He-Ne laser, and the beam is divided into transmission beam and reflection beam by the beam splitting mirror.  Each beam is reflected by the deflection mirror and the flat mirror respectively, and they converge on the ground glass screen to form circular interference fringes.

By rotating the spiral micrometer to adjust the forward and backward position of the flat mirror, the thickness and density of the interference fringes are controlled. Simultaneously, adjusting the pitch and yaw angles of the flat mirror changes the position of the interference ring center on the screen. The adjusted interference pattern will appear in the center of the ground glass screen, and a clear interference pattern will also be present on the other side of the screen. The CCD camera behind the ground glass screen captures black-and-white photographs of the interference pattern, which are then sent back to the computer for processing.

In the experiment, the deflection mirror undergoes deflection by adjusting the voltage applied to it. After deflection, the deflection mirror performs self-testing and returns a deflection angle. Before the experiment, the returned values are calibrated using an automatic collimator with a resolution of 0.02", model CONEX-LDS. The measurement range is from 4.12" to 206.26" with a step size of 4.12". The results are shown in Fig 3.2, which validates the accuracy of the returned values.

\begin{figure}[htbp]
\renewcommand{\thefigure}{3.2}
\centering
\subfigure[]{
\label{time.1}
\includegraphics[width=5cm]{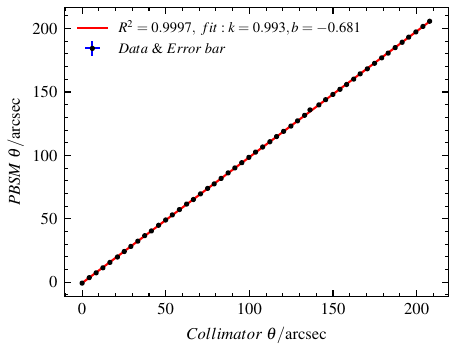}
}
\subfigure[]{
\label{time.2}
\includegraphics[width=5.1cm]{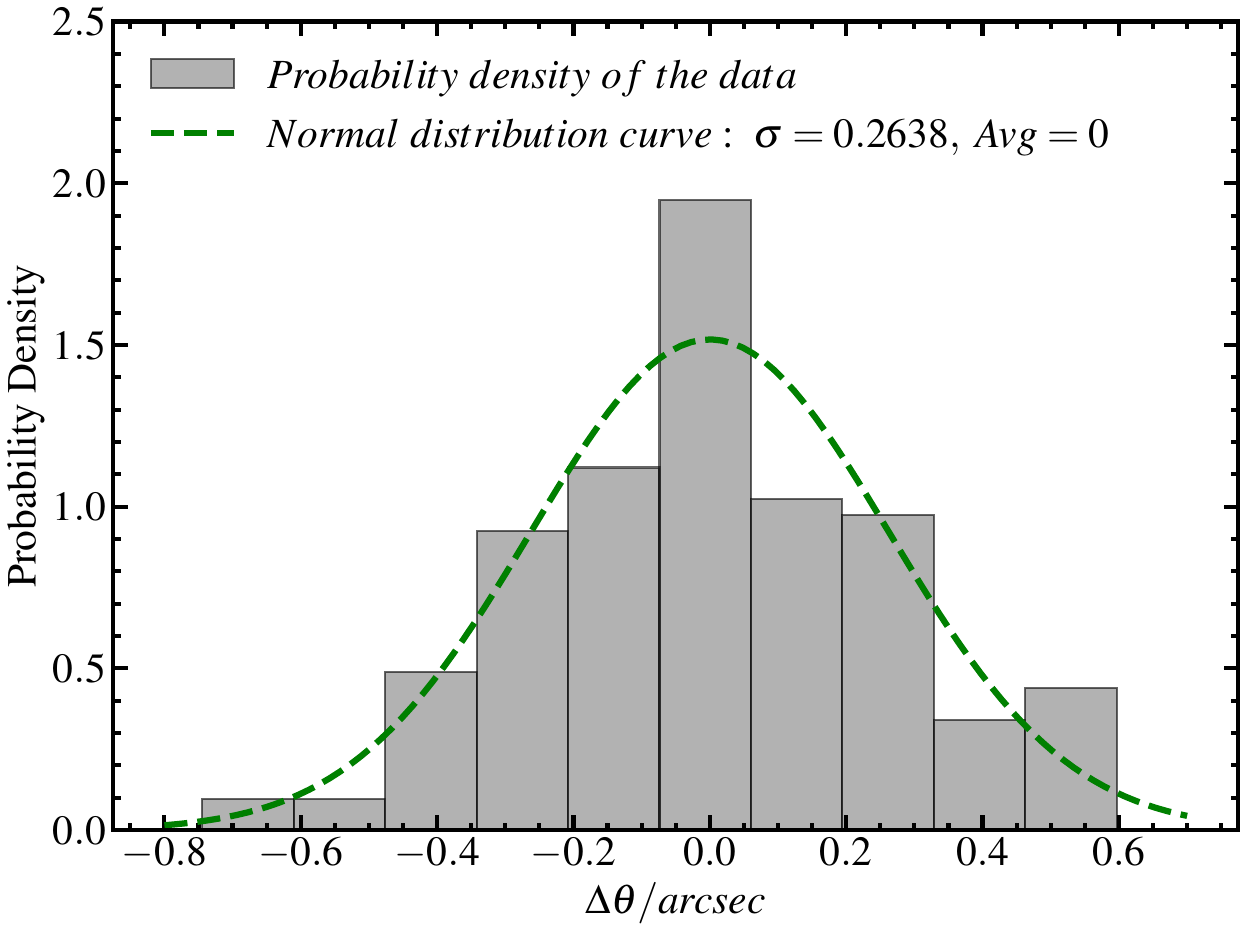}
}
\caption{(a) shows the relationship between the measurements obtained from the automatic collimator and the values returned by the deflection mirror, fitted using the function $y=kx+b$.In the experiment, all returned values are corrected using this relationship equation before being used as true values. (b) shows the difference between the corrected returned values and the collimator measurement values. The average error is $7.40\times10^{-15}$", and the standard deviation is $0.264$".}
\label{图1的标签}
\end{figure}

\subsection{Experimental Process}
In the experiment, initially, stable interference fringes are adjusted to appear at the center of the screen. Then, by rotating the spiral micrometer, the distance between the flat mirror and the beam splitter is adjusted so that the number of interference rings reduces from 1 to 3, with the fringe order kept as low as possible. At this point, the optical path lengths of the two paths cannot yet be equal, so fine adjustment is needed using interference from a light source with a shorter coherence length.For fine-tuning, a mercury lamp with a ground glass diffuser is used to replace the He-Ne laser, and the ground glass receiver screen is replaced with a mirror. Interference fringes from the mercury lamp are observed in the mirror. Finally, by rotating the spiral micrometer, the optimal number of interference rings in the field of view is adjusted to 1 or 2, and zero-order interference fringes are observed, indicating the completion of the equalization of the arm lengths.The reading of the spiral micrometer is recorded, and subsequently, the micrometer can be used directly to measure the arm length difference. Through this method, theoretically, the measurement of the arm length difference can be accurate to 0.01mm. It will be explained later in the text that precise measurement of the arm lengths is required by the geometric method.
~\\

\begin{figure}[htbp]
\renewcommand{\thefigure}{3.3}
\centering\includegraphics[width=10cm]{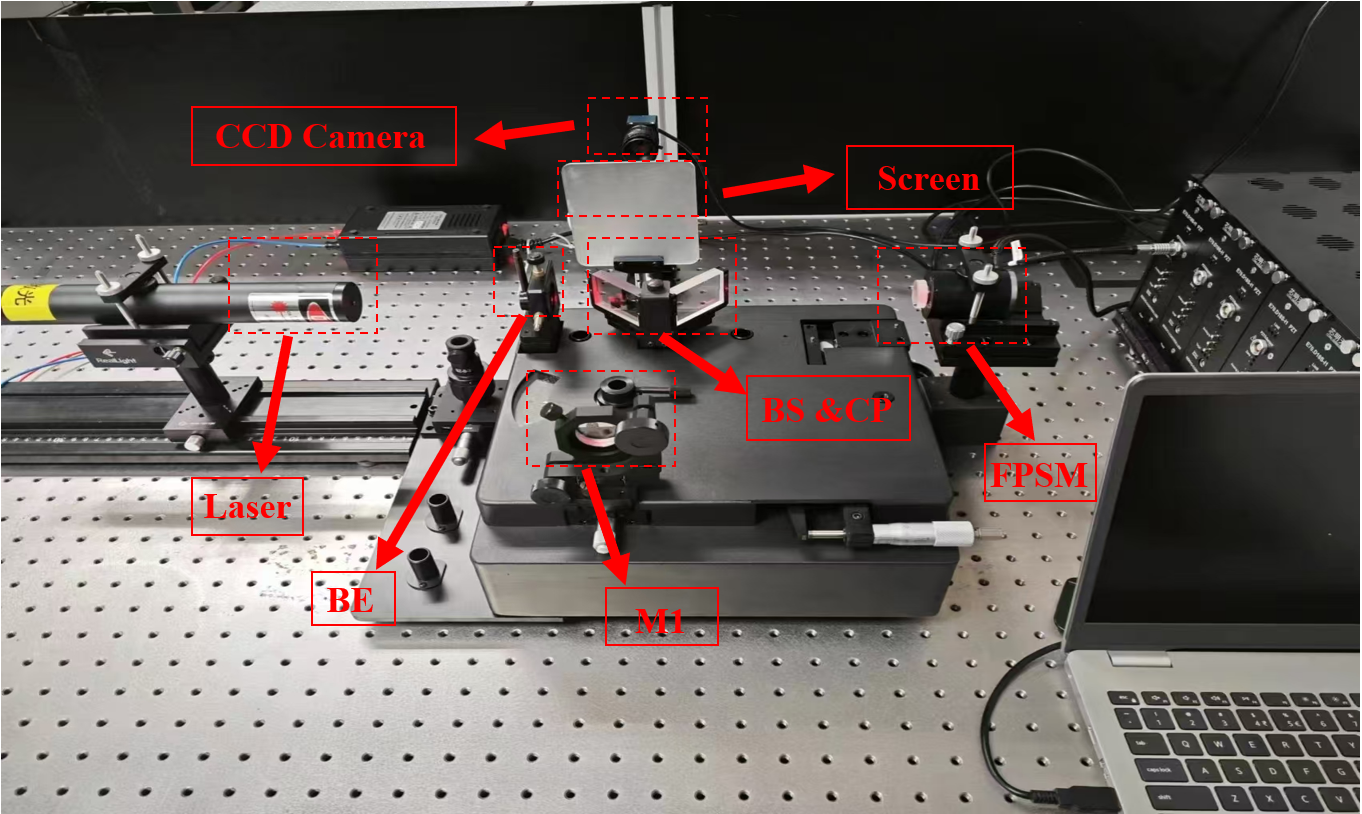}
\caption{The diagram of the experimental platform, with the placement of each component labeled as shown in the figure, includes the following elements: Laser is a 632.8nm He-Ne laser, BE is a beam expander with a focal length of 6.2mm, BS is the beam splitter, CP is the compensating plate, CCD has a resolution of 1280×1080 pixels, and POM is the deflection mirror.}
\end{figure}

Subsequently, the He-Ne laser and the ground glass screen are reinstalled. The spiral micrometer is adjusted so that there are approximately 10 equal inclination interference rings. The initial deflection angle of the deflection mirror is set to 0, and the initial interference image on the ground glass is recorded by using the CCD.Next, the deflection mirror is controlled by voltage to produce small incremental deflection angles with a step size of $4.12$", causing the equal inclination interference rings to move on the screen. The CCD records the interference image for each displacement. After completing the measurements, the size of the CCD pixels is calibrated, and a laser rangefinder with an accuracy of 2mm is used to measure the length of each arm.

The data processing in the experiment is shown in fig 3.4.
\begin{figure}[htbp]
\renewcommand{\thefigure}{3.4}
\centering
\subfigure[]{
\label{time.1}
\includegraphics[width=3.8cm]{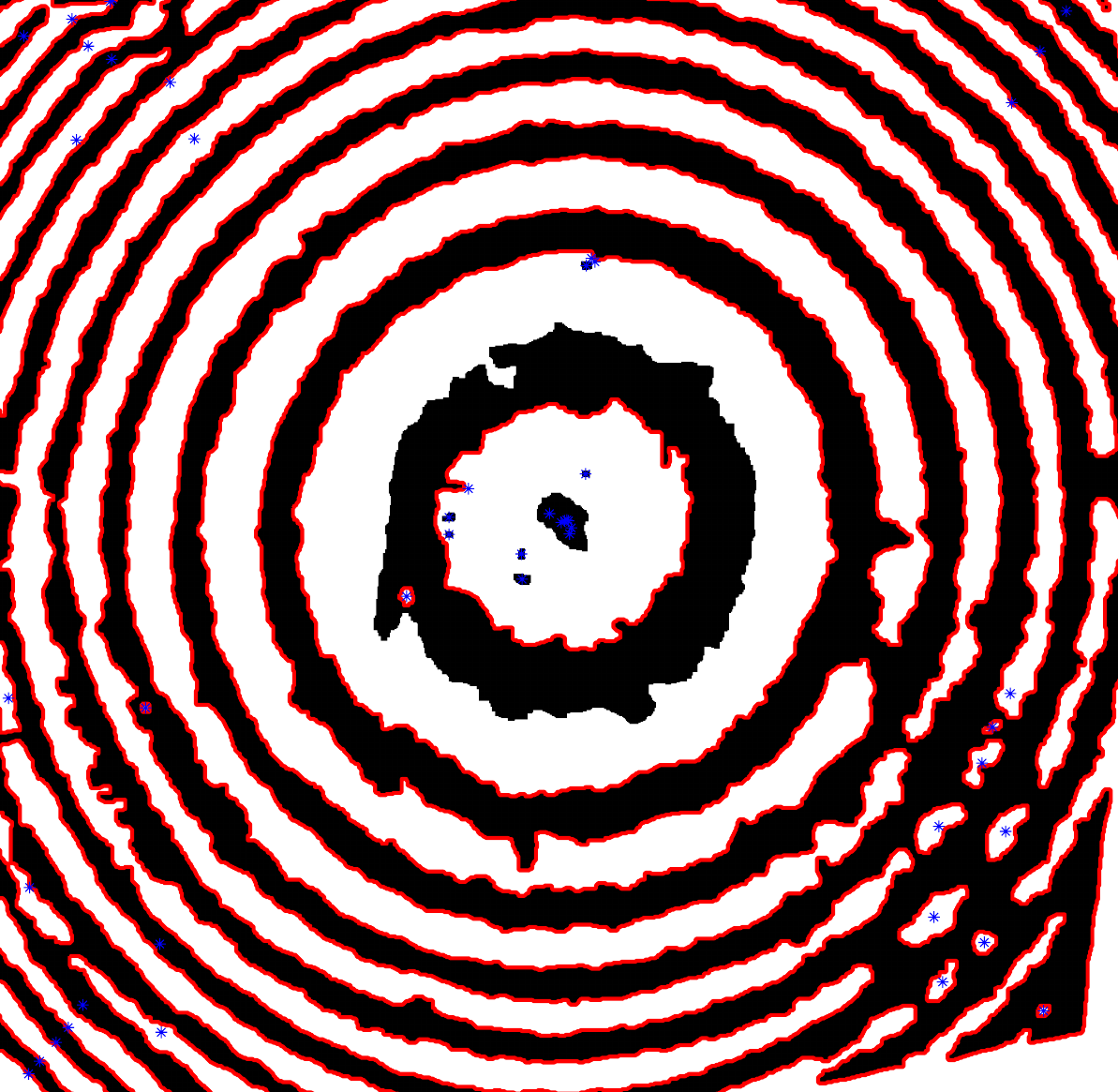}
}
\subfigure[]{
\label{time.2}
\includegraphics[width=3.5cm]{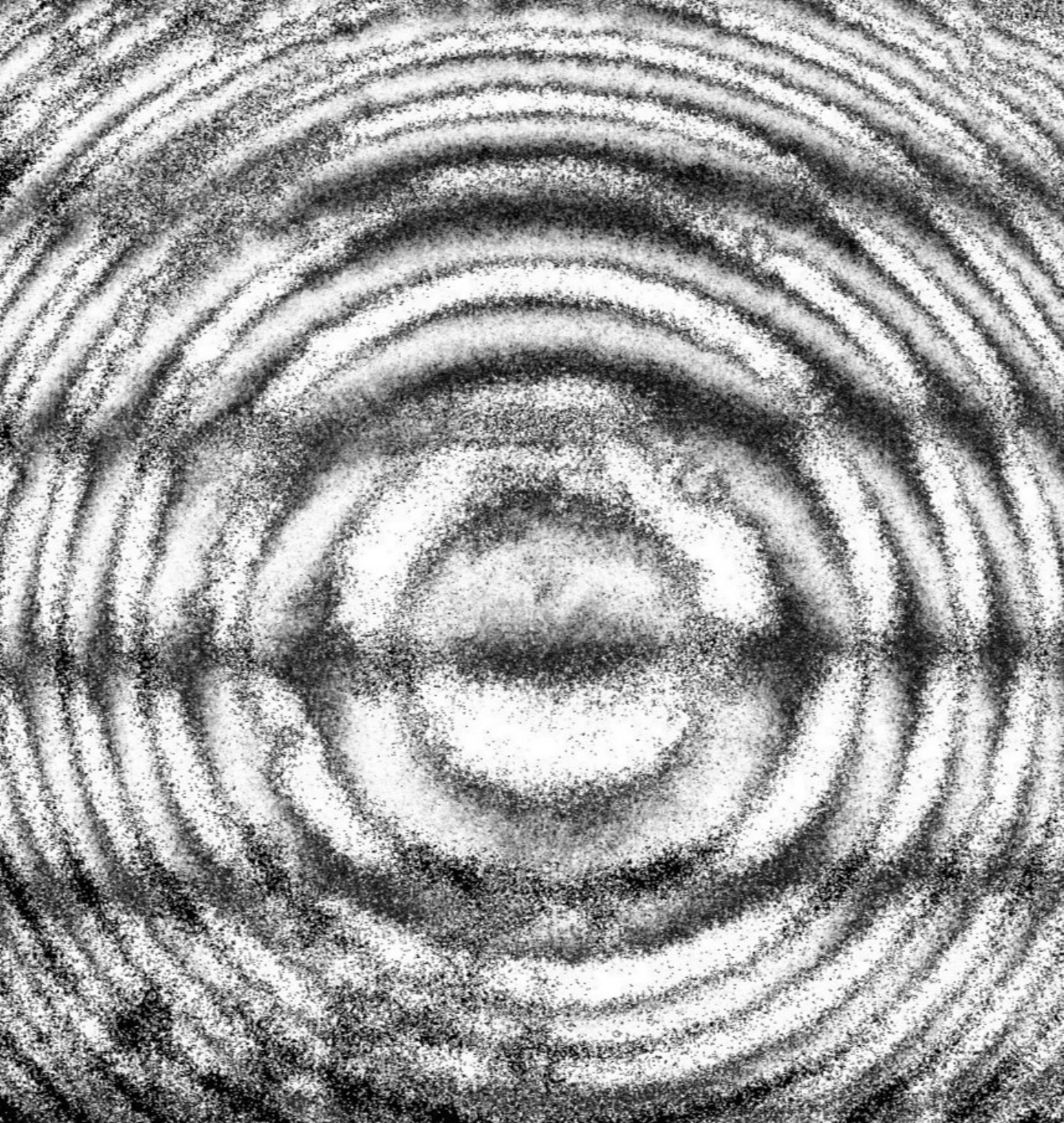}
}
\subfigure[]{
\label{time.3}
\includegraphics[width=3.5cm]{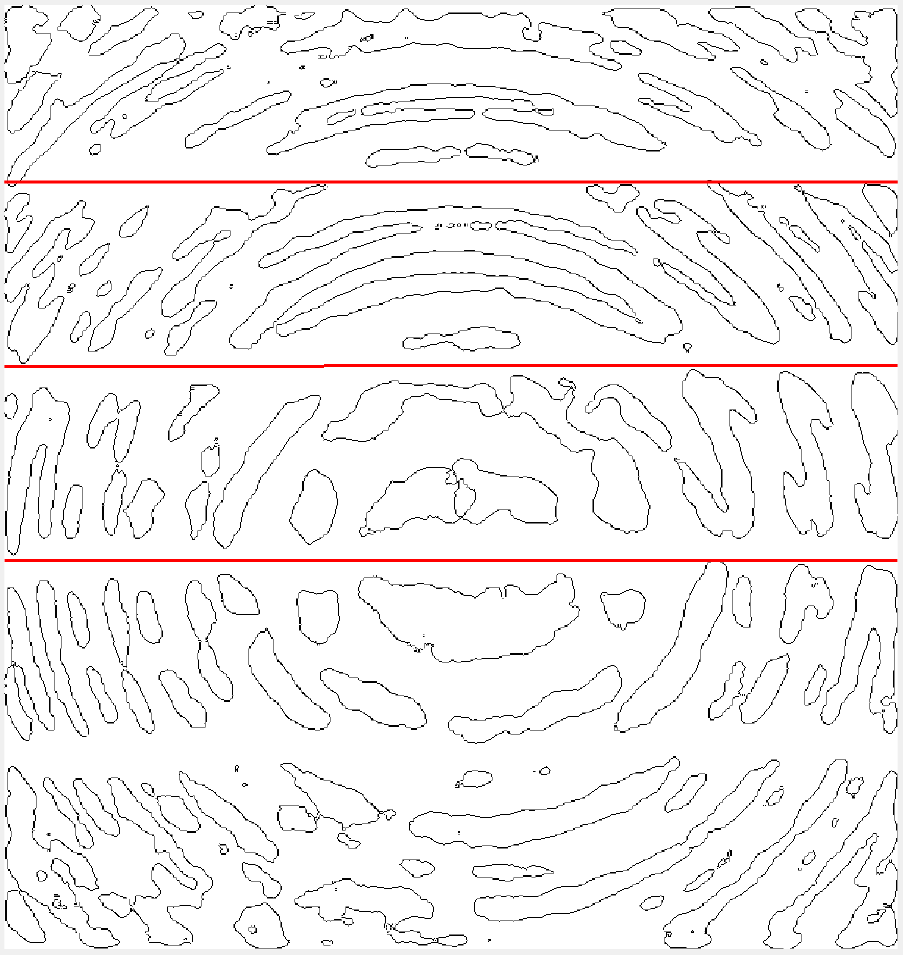}
}
\caption{(a)shows the extraction of the center coordinates using the geometric method;(b)presents the image processed by the ESPI program;(c)displays the result of edge extraction after binarizing Figure 2.}
\label{fig:图像识别}
\end{figure}

\subsection{Results}
Fig 3.5 presents the experimental results obtained from processing the same set of experimental data using both the ESPI method and the geometric method. In the experiment, the deflection angle of the deflection mirror varied from 12.37" to 45.24", changing by 4.12" each time, with three repeated measurements taken. After processing the data, the average values of the three sets of experimental results were calculated, yielding $\theta_{ESPI}$ (ESPI method) and $\theta_{Geo}$ (geometric method).
\begin{figure}[htbp]
\renewcommand{\thefigure}{3.5}
\centering
\subfigure[]{
\label{time.1}
\includegraphics[width=6cm]{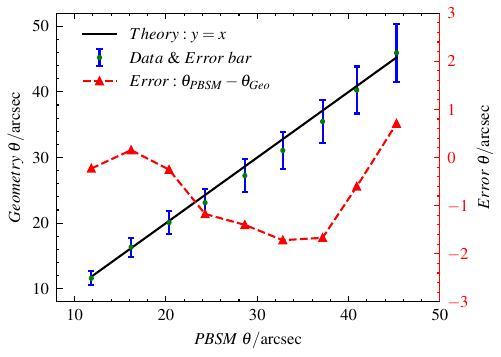}
}
\subfigure[]{
\label{time.2}
\includegraphics[width=6cm]{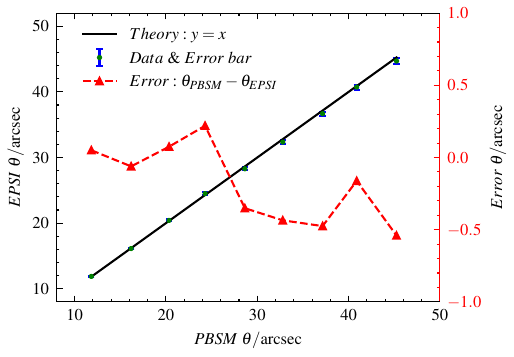}
}
\caption{(a) The experimental results of the geometric method: the black solid line represents the theoretical curve, the green scatter points represent the processed angle measurement results, the blue bars represent the error bars, and the red dashed line represents the absolute error of each measurement point.(b) The experimental results of the ESPI method: the points and lines in the image have the same meanings as those in (a).}
\label{图1的标签}
\end{figure}

Comparing the deflection mirror's recorded return values $\theta$ with $\theta_{ESPI}$ and $\theta_{Geo}$ reveals the measurement errors of these two methods. The experimental results show that the ESPI method has a maximum relative error of -1.33\%, a minimum relative error of -0.37\%, and an overall average error of 0.16\%. The geometric method has a maximum relative error of -6.26\%, a minimum relative error of -0.97\%, and an overall average error of 3.35\%.

The experimental results demonstrate the accuracy of the ESPI method and the geometric method in measuring small angles. Based on the experimental results, further analysis of the experiment and theory is conducted regarding accuracy and resolution.
For the geometric method and the ESPI method, the accuracy of angle measurement is evaluated by uncertainty. The uncertainty considers two components: the type A uncertainty, represented by the standard deviation of the mean of three repeated measurements, and the type B uncertainty, which includes factors such as the systematic errors of the experimental apparatus. The combined expanded uncertainty represents the precision of the experiment.The type B uncertainty in the measurement results primarily considers the measurement of the optical arm lengths, calibration of the CCD pixel size, jitter in the deflection mirror's returned values, errors in the extraction algorithm, and jitter in the interference rings. The calculated results are shown in Fig 3.6. The uncertainty for the geometric method ranges from 1.0" to 4.4", while the uncertainty for the ESPI method ranges from 0.095" to 0.43".
% \vspace{2em}

The experimental results indicate that the accuracy of the ESPI method is superior to that of the geometric method, which is consistent with theoretical expectations. Comparing the formulas for the geometric method (Eqs.\eqref{2}) and the ESPI method (Eqs.\eqref{8}), the denominator in Eqs. \eqref{2} is the optical arm difference $2d_2-2d_3$, which makes the geometric method more sensitive to errors in the measurement of optical arm lengths. Adjusting the lengths of $d_2$ and $d_3$ before the experiment is done to use a spiral micrometer to measure $d_2-d_3$ and reduce measurement errors.Furthermore, the uncertainty of the geometric method exhibits a certain linear relationship with the deflection angle, while the ESPI method does not. This is because the error of the geometric method mainly depends on the measurement error of the optical arm lengths, leading to non-statistical analysis dominance of type B uncertainty.
~\\

\begin{figure}[htbp]
\renewcommand{\thefigure}{3.6}
\centering
\includegraphics[width=7cm]{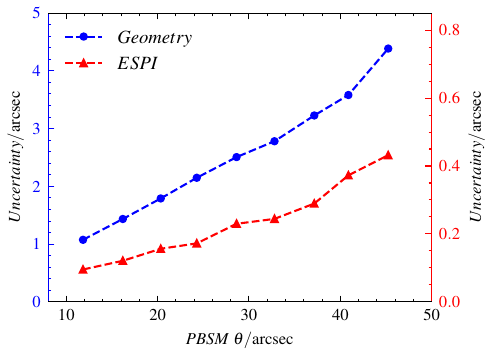}
\caption{The calculation results of the uncertainty for the geometric method and the ESPI method are shown in the figure, with the blue representing the precision of the geometric method and the red representing the precision of the ESPI method.}
\end{figure}

As the deflection angle of the measurement increases, the uncertainty of both the geometric method and the ESPI method also increases, as shown in Fig 3.6. This indicates a decrease in the precision of both angle measurement methods. This is because as the deflection angle increases, the higher-order errors under the first-order approximation of Eqs.\eqref{2} and \eqref{8} become significant. Additionally, the interference fringe spacing decreases for the ESPI method, and the distance of the centroid of the geometric method from its initial position increases. As a result, the error propagation coefficient increases, leading to a decrease in precision.

\begin{figure}[htbp]
\renewcommand{\thefigure}{3.7}
\centering
\subfigure[]{
\label{time.1}
\includegraphics[width=7.8cm]{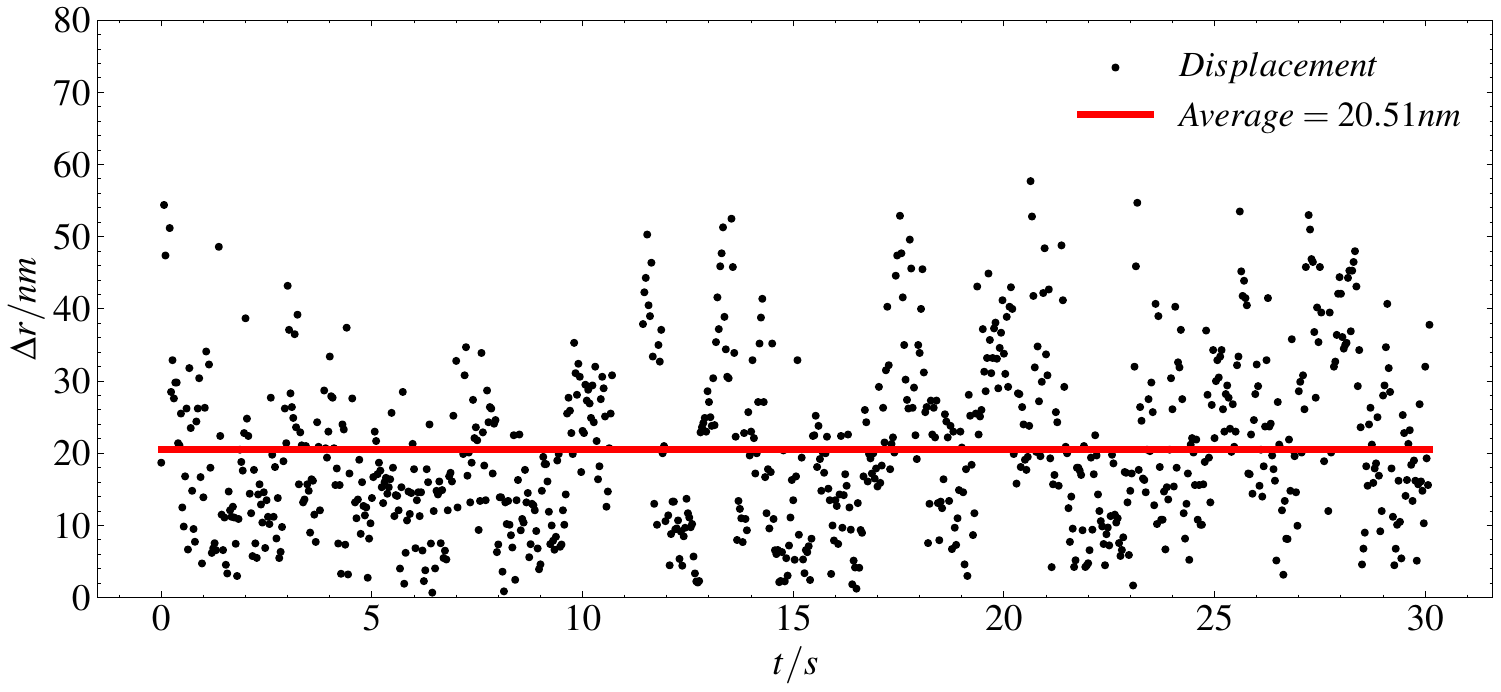}
}
\subfigure[]{
\label{time.2}
\includegraphics[width=4.8cm]{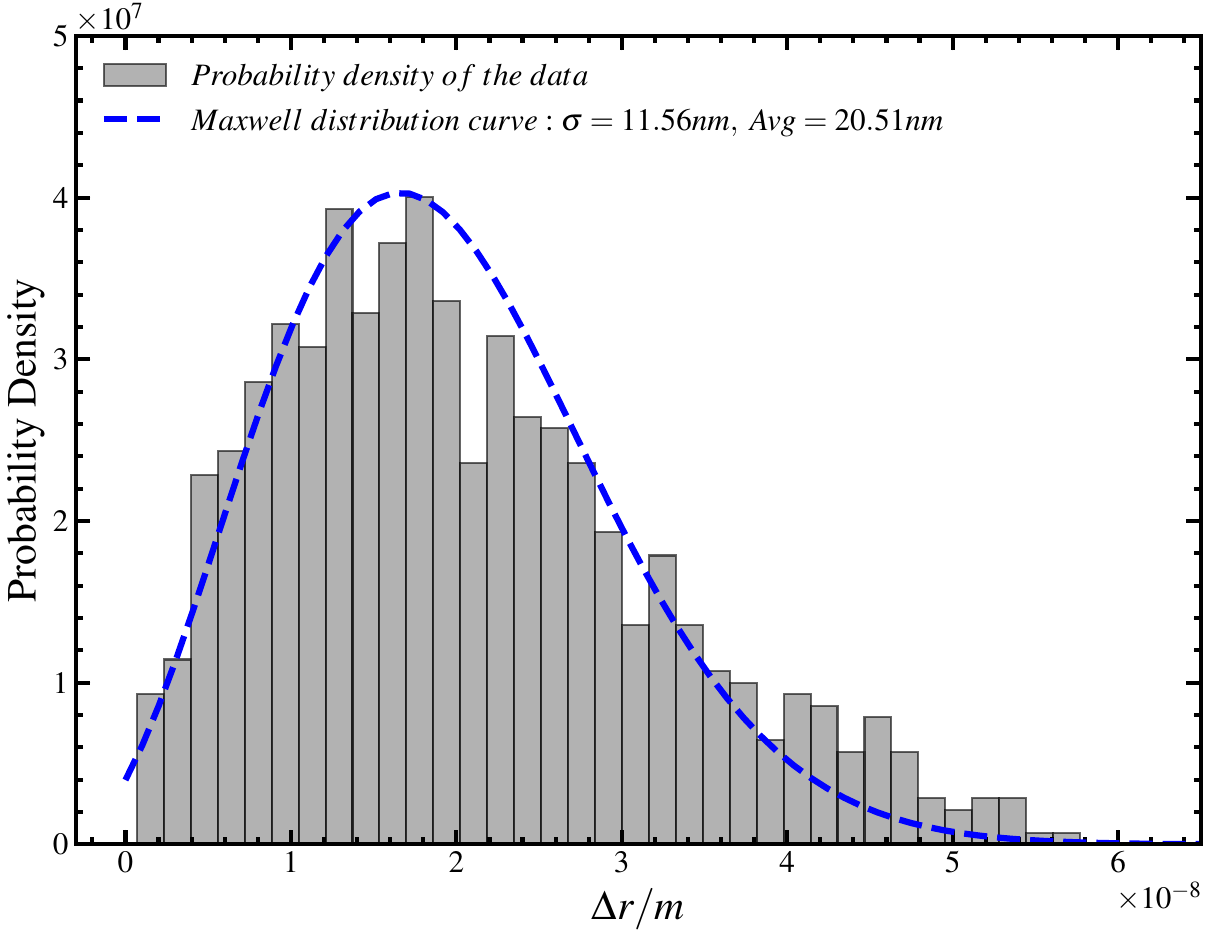}
}
\caption{(a) The recorded results of laser jitter within 30 seconds after excluding gross errors. The data were recorded 24 times per second, and the red line represents the average jitter.(b) Histogram of laser jitter, where the jitter displacement $\Delta r$ is calculated as $\sqrt{\Delta x^2+\Delta y^2}$, with $x$ and $y$ representing the coordinates of the centroid positions of the spots on the screen. Since $\Delta x$ and $\Delta y$ follow a normal distribution, $\Delta r$ follows a Maxwell distribution. From (b), it can be observed that the histogram distribution closely matches the Maxwell distribution (blue dashed line).}
\label{图1的标签}
\end{figure}

For the analysis of resolution, the resolution of the ESPI method mainly depends on the measurement of fringe spacing and laser spatial jitter. The standard deviation of the measured laser spatial jitter is $11.56$ nm (Fig 3.7), which corresponds to $0.018$" in the experiment. Considering error propagation, the maximum standard deviation of the ESPI method measurement is $0.266$". The laser jitter corresponds to a small angle of $\Delta \theta = \frac{\Delta \lambda(d_1+2d_2+d_4)}{2\Delta x (d_1+d_2) } =0.018$". Using one standard deviation ($\sigma$) as the criterion for resolution, the theoretical resolution of the ESPI method is determined to be $0.284$". The resolution of the geometric method depends on the measurement of circle center positions and laser spatial jitter. Considering error propagation, the maximum standard deviation of the geometric method angle measurement is $1.81$". According to the one-$\sigma$ principle, the resolution of the geometric method is calculated to be $1.828$".

~\\

%In summary, the experiment verifies that within the measurement range of 0.015°, the ESPI method achieves an accuracy of less than $0.43$" and a resolution of $0.284$", while achieving an accuracy of $0.095$" for angle measurements below $16$". For the geometric method, an accuracy of less than $4.39$" and a resolution of $1.828$" are obtained, with an accuracy of $0.716$" achievable for angle measurements below $10$". This demonstrates the high precision and high resolution capabilities of both methods.\\

%\subsection{几何法测量范围的仿真分析}%

%可能要移到前面的仿真部分？%

\section{Conclusion}
This paper utilizes an ESPI method based on a Michelson interferometer to implement different angle measurement methods within various angle ranges. Two exposures record the changes in the light field distribution before and after the angle adjustment, and relevant algorithms are designed to calculate the centroid movement distance and fringe spacing required by both methods. The deflection angle is then deduced from these formulas. Theoretically, it has a large angle measurement range of 0.6° while maintaining high precision and resolution. Experimental results show that for angle measurements not exceeding 54", the ESPI method can achieve an accuracy of less than 0.43" and a resolution of 0.284". When measuring angles less than 16", the accuracy of the ESPI method can be improved to 0.095". In contrast, the geometric method offers an accuracy of less than 4.39" and a resolution of 1.828" within the same measurement range. When measuring angles less than 10", the geometric method's accuracy can improve to 0.716", still providing high precision and resolution. The method proposed in this paper is structurally simple and capable of real-time high-precision measurement of deflection angles, demonstrating robust performance. It can be effectively applied in precision manufacturing, aerospace, and various scientific research tests.
%\newpage
%%%%%%%%%%%
\bibliography{reference}
%%%% bib file name
%%%%%%%%%%%

\end{document}